\documentclass[aps,prc,a4paper,nofootinbib,
preprintnumbers,superscriptaddress,showpacs,showkeys,twocolumn]{revtex4-2}

\usepackage{color}
\usepackage{soul,color}
\usepackage{graphicx}
\usepackage{epsfig}
\usepackage{bm}
\usepackage{amsmath}
\usepackage{amssymb}
\usepackage{xcolor}
\usepackage[colorlinks,
linkcolor=red,
anchorcolor=red,
urlcolor=red,
citecolor=red,
]{hyperref}

\newcommand{\be}{\begin{equation}}
\newcommand{\ee}{\end{equation}}
\newcommand{\ba}{\begin{eqnarray}}
\newcommand{\ea}{\end{eqnarray}}

\begin{document}
\title{Heavy quarks in the early stage of high energy nuclear collisions at RHIC and LHC: 
Brownian motion versus diffusion in the evolving Glasma} 

\author{Pooja}
\affiliation{School of Physical Sciences, Indian Institute of Technology Goa, Ponda-403401, Goa, India}

\author{Santosh K. Das}
\affiliation{School of Physical Sciences, Indian Institute of Technology Goa, Ponda-403401, Goa, India}

\author{Lucia Oliva}
\affiliation{Department of Physics and Astronomy, University of Catania,
	Via S. Sofia 64, I-95123 Catania, Italy}

\author{Marco Ruggieri}\email{
	ruggieri@lzu.edu.cn}
\affiliation{School of Nuclear Science and Technology, Lanzhou University, 222 South Tianshui Road, Lanzhou 730000, China}

\begin{abstract}

We study the transverse momentum, $p_T$,
 broadening of charm and beauty quarks
in the early stage of high energy nuclear collisions.
We aim to compare the diffusion in the evolving Glasma fields
with that of a standard, Markovian-Brownian motion
in a thermalized medium with the same energy density
of the Glasma fields. 
The Brownian motion is
studied via Langevin equations with diffusion coefficients
computed within perturbative Quantum Chromodynamics.
We find that for small values of the saturation scale,
$Q_s$, the average $p_T$ broadening in the two motions is 
comparable, that suggests that the diffusion coefficient 
in the evolving Glasma fields is in agreement with the
perturbative calculations. On the other hand, for large $Q_s$ 
the $p_T$ broadening in the Langevin motion is smaller than that
in the Glasma fields. This difference can be
related to the fact that heavy quarks in the latter system
experience diffusion in strong, coherent gluon fields
that leads to a faster momentum broadening due to memory,
or equivalently to a strong correlation in the transverse plane.

\vspace{2mm}
\noindent {\bf PACS}: 25.75.-q; 24.85.+p; 05.20.Dd; 12.38.Mh

\noindent{\bf Keywords}:Relativistic heavy-ion collisions, heavy quarks, Glasma, Classical Yang-Mills equations, Wong equations, Langevin equations, Brownian motion, memory effects.
\end{abstract}
\maketitle

\section{Introduction}

The research of the pre-thermalization phase of the system formed in high-energy collisions 
at the Relativistic Heavy Ion Collider (RHIC) and at the
Large Hadron Collider (LHC),
and of its evolution to the Quark-Gluon Plasma (QGP)~\cite{Shuryak:2004cy,Jacak:2012dx} is a fascinating topic in the field of heavy-ion phenomenology. 
The high energy collisions are modeled as those 
of two sheets of colored glass~\cite{McLerran:1993ni,McLerran:1993ka,McLerran:1994vd,Gelis:2010nm,Iancu:2003xm,McLerran:2008es,Gelis:2012ri}. 
According to the effective field theory of color-glass condensate (CGC), 
the interaction of two colored glasses 
generates strong longitudinal 
gluonic fields in the forward light cone. 
These fields are called the Glasma~\cite{Kovner:1995ja,Kovner:1995ts,Gyulassy:1997vt,Lappi:2006fp, Krasnitz:2003jw,Fukushima:2006ax,Fujii:2008km,Fukushima:2013dma,Romatschke:2005pm, Romatschke:2006nk,Fukushima:2011nq}. 
Glasma is a set of classical fields due to their large
intensity, $A_\mu^a \simeq Q_s/g$, where 
$Q_s$ is the saturation scale and
$g$ is the
coupling of Quantum Chromodynamics (QCD). 
The time evolution of these gluon fields is described by Classical Yang-Mills (CYM) equations. In this work, we reserve the notation EvGlasma for the evolving Glasma 
fields and preserve the name Glasma for the initial condition.
The duration of the EvGlasma is that of the
pre-hydro stage, therefore  typical duration of this stage
is of the order of $1~\mathrm{fm/c}$ for AA collisions
at the RHIC energy, and approximately $0.3$ fm/c for AA
collisions at LHC.

Heavy quarks~\cite{Prino:2016cni, Andronic:2015wma, Rapp:2018qla, Cao:2018ews, Aarts:2016hap, Greco:2017rro, Dong:2019unq, Xu:2018gux, Moore:2004tg, vanHees:2005wb, vanHees:2007me, Gossiaux:2008jv, He:2011qa, Prakash:2021lwt, Song:2015sfa, Alberico:2011zy, Lang:2012cx, Xu:2017obm, Cao:2016gvr, Das:2016cwd, Das:2015ana, Das:2017dsh, Das:2015aga,Song:2019cqz, Beraudo:2015wsd,Das:2013kea, Scardina:2017ipo, Mrowczynski:2017kso} are produced in the primordial 
stage of the high energy nuclear collisions are noble 
probes of EvGlasma. 
In fact, their production time is approximately 
$\tau_{form} \approx 1/(2m)$, where $m$ is the mass of heavy quark. Thus, heavy quarks are produced shortly after the collision
and potentially diffuse through the medium, testing 
in particular
the pre-hydro stage of the system. 

The diffusion of heavy quarks
in the EvGlasma can be different from that in a
thermal medium with uncorrelated noise. In fact,
in the former case the heavy quarks diffuse among
domains in which the fields are correlated,
that have transverse area $\approx 1/Q_s^2$,
namely the flux tubes or the filaments.
The arrangement of the filaments in the transverse plane
is random, therefore if heavy quarks are slow enough, they
can diffuse within a single filament for a finite amount of time
and experience the coherent gluon field. Instead, if the transverse
velocity of the heavy quark is large enough, it effectively
jumps between uncorrelated filaments and its diffusion is more
similar to that in a random medium.
A corollary of this qualitative picture is that heavy quarks
with larger mass, namely beauty quarks, experience the strong
coherent gluonic field mode than the charm quarks. This is
confirmed by our calculations, see Section III for details.

It is of a certain interest to compare the motion of heavy colored probes in the EvGlasma,
with that in a hot thermalized medium. 
This is the main subject of our study.
From previous studies \cite{ Ruggieri:2018rzi,Sun:2019fud,Liu:2019lac,Boguslavski:2020tqz}
it is known that the evolution of heavy quarks in the strong gluon fields produced in the early stage of high energy nuclear collisions
is even qualitatively different from a standard Brownian motion, due to memory effects that make the diffusion ballistic
\cite{Liu:2020cpj}. In fact, as a result of memory in the gluon fields, the transverse momentum broadening in the evolving Glasma
overshoots the one that the heavy quarks would experience in a hot QCD plasma.
It is useful to remark that this memory effect in the early stage is important due to 
the fact that the memory time, $\tau_\mathrm{mem}\approx 0.1$ fm/c
is comparable with the lifetime of the early stage, 
as well as to the fact that thermalization time $\tau_\mathrm{therm}\gg\tau_\mathrm{early}$:
the combination of these makes the perfect scenario for observing memory effects,
as well as to make the early evolution of heavy quarks resemble a diffusive motion with negligible drag.

In particular, we perform a systematic comparison of 
the diffusion in the EvGlasma and in a hot medium,
computing the transverse momentum, $p_T$, broadening 
	\begin{equation}
	\sigma_p = \frac{1}{2} \langle (p_x(t)-p_{0x})^2 + (p_y(t)-p_{0y})^2 \rangle,
	\label{e32}
\end{equation}
where $p_{0x}$ and $p_{0y}$ are the initial $x$ and $y$ components of the initial transverse momentum. Hence, $p_T^2 = p_x^2 + p_y^2$.
In realistic collisions the geometry of the early stage is
that of a longitudinally expanding medium.
However,
we firstly analyze the diffusion in a static box. 
The use of the static box is necessary
because the estimate and the comparison 
of the diffusion coefficients
of the heavy quarks in the two systems should be done
firstly for configurations with a fixed energy density,
or equivalently with a fixed temperature. After this scenario
is clarified, it is possible to elucubrate on similarities
and differences of the diffusion in systems with expansion.
Therefore, the use of a static box is far from being academic:
it is the natural framework to define the diffusion coefficients.
In the EvGlasma case we find that $\sigma_p\propto t^2$ in the
very early stage, while $\sigma_p\propto t$ in the later
stage in analogy with the standard Brownian motion.
This behavior, that was anticipated in \cite{Liu:2020cpj},
leads to a fast increase of the $\sigma_p$,
in fact faster than that found within Langevin equation
with the pQCD diffusion coefficient.

We analyze briefly the role of the longitudinal expansion
on $p_T$ broadening in the EvGlasma fields. We show that
in this case, $\sigma_p$ tends to saturate at the end
of the EvGlasma phase. This late time 
behavior looks similar to that
of the standard Brownian motion and in that context
it is understood as the equilibration of the 
heavy quarks with the medium: 
however, the reason why
$\sigma_p$ saturates in the EvGlasma is not related
to the equilibration of the heavy quarks, rather to the
dilution of the energy density during the expansion.

The implications of this early stage scenario are far from being purely abstract. As a matter of fact,
it has been shown that the diffusion in the evolving Glasma can affect observables,
namely the nuclear modification factor, $R_\mathrm{AA}$, the elliptic flow and the momentum broadening parameter
of hadrons containing heavy quarks  
\cite{Ruggieri:2018rzi,Sun:2019fud,Ipp:2020nfu}.

The plan of the article is as follows. 
In Section II, we review the formalism of the work.  
In Section III, we present our results on $\sigma_p$
for the EvGlasma, the thermalized medium and the EvGlasma
with longitudinal expansion. Finally, in Section IV we present
our conclusions and discuss possible improvements that will
be implemented in future works.

\section{Formalism}
 \subsection{Glasma and classical Yang-Mills equations for its evolution}
We summarize here the McLerran-Venugopalan (MV) model. 
According to the effective field theory of color-glass condensate (CGC), the two colliding objects can be described as two thin Lorentz-contracted sheets of a colored glass where fast parton dynamics is frozen by time dilation. These fast partons act as the sources for the slow gluon fields in the saturation regime and thus behave like classical fields. 

In the MV model, the sources of the CGC fields are the fast partons which are represented by the randomly distributed static color charge densities in the two colliding nuclei,
that we label as $A$ and $B$.
The static color charge densities $\rho^a_{A}$ on the colliding nucleus $A$ are considered as random variables which are normally distributed with zero mean and variance specified by 
\begin{equation}
	\langle \rho^a_{A}(\textbf{\textit{x}}_T)\rho^b_{A}(\textbf{\textit{y}}_T)\rangle=  (g\mu_A)^2  \delta^{ab} \delta^{(2)}(\textbf{\textit{x}}_T - \textbf{\textit{y}}_T).
	\label{eq:colorchag88} 	
\end{equation}
Here, $a$ and $b$ indicate the adjoint color index. This work is limited to the $SU(2)$ color group, hence $a, b=1,2,3$. 
$\mu$ is the density of the color charge carriers in the
transverse plane, therefore $g\mu$ 
denotes the color charge density; $g^2\mu = O(Q_s)$ \cite{Lappi:2007ku}
where $Q_s$ is the saturation momentum.

The static color sources ${\rho}$ generate pure gauge fields outside and on the light cone. They combine in the forward light cone and give the initial boost-invariant Glasma fields. To determine these fields, firstly we solve
the Poisson’s equations for the gauge potentials generated by the color charge distribution of the $i^{th}$ nuclei,
\begin{equation}
	-\partial^2_{\perp}\Lambda^{(i)}(\textbf{\textit{x}}_T)
	=\rho^{(i)}(\textbf{\textit{x}}_T),
	\end{equation}
where $i=A, B$. 
The Wilson lines are calculated as $V^\dagger(\textbf{\textit{x}}_T) =e^{-ig\Lambda^{(A)}(\textbf{\textit{x}}_T)}$ and $W^\dagger(\textbf{\textit{x}}_T) =e^{-ig\Lambda^{(B)}(\textbf{\textit{x}}_T)}$
and the gauge fields of the two colliding objects are given as
\begin{eqnarray}
  \alpha_i^{(A)} &=& \frac{-i}{g}V\partial_iV^\dagger,
  \label{eq:alpha1pp}\\
  \alpha_i^{(B)} &=& \frac{-i}{g}
  W\partial_iW^\dagger.\label{eq:alpha2pp}
\end{eqnarray}
The solution of the CYM in the forward light cone at initial time, namely, the Glasma gauge potential in terms of the gauge fields can be written as
\begin{equation}
    A_i = \alpha_i^{(A)} +\alpha_i^{(B)},~~~A_z=0,
\end{equation}
where $i = x, y$. Assuming $z$ to be the direction of the collision, the initial longitudinal Glasma fields are
\begin{eqnarray}
	E^z &=& -ig \sum_{i=x,y}\bigg[\alpha^{(B)}_i, \alpha^{(A)}_i\bigg],\\
	B^z &=& -ig \bigg(\bigg[\alpha^{(B)}_x, \alpha^{(A)}_y\bigg]+\bigg[\alpha^{(A)}_x, \alpha^{(B)}_y\bigg]\bigg),
\end{eqnarray}	
while the transverse fields vanish at the initial time.

The evolution of the pre-equilibrium period of the heavy-ion collisions, namely the Glasma fields is described by the classical Yang-Mills (CYM) equations. We look after the QCD evolution of the dense gluonic background fields at the classical level using CYM equations which are nothing just the non-abelian generalization of the well known Maxwell equations in the empty space. The equations of the motion of the fields and the conjugate momenta, i.e., the CYM equations are
\begin{eqnarray}
&&	\frac{dA^a_i(x)}{dt}= E^a_i(x),\\
&&	\frac{dE^a_i(x)}{dt}= \partial_jF^a_{ji}(x)+g f^{abc} A^b_j(x)F^c_{ji}(x),
\end{eqnarray}	
 where $f^{abc} = \varepsilon^{abc}$ with $ \varepsilon^{123}=+1$ and $F^a_{ij}$ is the magnetic part of the field strength tensor given by
 \begin{equation}
	F^a_{ij}(x) = \partial_i A^a_j(x)- \partial_j A^a_i(x)+ g f^{abc} A^b_i(x)A^c_j(x)
\end{equation}
Here, we have used the standard Einstein summation convention. In this work, we do not include 
the color current carried by the heavy quarks: this would be
necessary to take the energy loss effects
into account, but
its effect has been studied previously and it has been shown that
it does not lead to any substantial effect within
the time range in which EvGlasma is physically relevant
\cite{Liu:2020cpj}.

\subsection{Wong equations for the heavy quarks} 

  Heavy quarks are created in the initial phase of the relativistic heavy-ion collisions in a very short time of 0.1 fm/c. The equations of motion of these colored probes in the Ev-Glasma are the Wong equations \cite{Liu:2019lac, Ipp:2020nfu, Ruggieri:2018rzi,Sun:2019fud, Wong:1970fu}

\begin{eqnarray}
	&&\frac{dx_i}{dt} = \frac{p_i}{E},\\
&&	\frac{dp_i}{dt}=g Q_a F^a_{i\nu}\frac{p^{\nu}}{E},\label{eq:wong2}
\end{eqnarray}
with $i=x,y,z$ and $E=\sqrt{\textbf{p}^2 + m^2}$ where $m$  
is the mass of the heavy quark.
These equations are the well known Hamilton equations of motion for the position and its conjugate momentum. 
In our setup,
the term on the right hand side of \eqref{eq:wong2} is 
the force that the EvGlasma fields exert on the heavy quarks.
Since the distribution of these fields is almost
random in the transverse plane, that force can be equated to
the random diffusive force in the Langevin equations,
see below. For the sake of nomenclature, 
we call this the Lorentz force,
because it is formally the nonabelian version of the 
force that the quarks would experience
in electromagnetic fields.

The 
third set of equations
describes the conservation
of the color current, namely
\begin{equation}
	\frac{dQ_a}{dt} = g \varepsilon^{abc}A^\mu_b   
	\frac{p_\mu}{E}Q_c,
\end{equation}
where $a=1, 2, \dots,N_c^2-1$ and $Q_a$ is the charm quarks effective color charge. Here, we initialize this color charge 
on a $3-$dimensional sphere with radius one; 
$Q_a Q_a$ corresponds to the total squared color charge
which is conserved in the evolution.
Differently from previous work, for example \cite{Liu:2020cpj},
in this work we use the QCD coupling, $g$, explicitly, 
to facilitate the comparison with the Langevin evolution
with pQCD diffusion coefficients.
For each heavy quark, we fix an anti-quark with the same initial position that of the companion quark but opposite momentum, 
while the color charge is chosen again randomly.

 \subsection{Brownian motion and Langevin dynamics}
The motion of a Brownian particle with mass $m$
is governed by 
the Langevin equations, namely
\begin{eqnarray}
 &&\frac{dx_i}{dt} = \frac{p_i}{E},
 	\label{l1}\\
 &&\frac{dp_i}{dt}= -\gamma p_i + \xi_i (t).
 \label{l2}
 \end{eqnarray}	
Here,  
$\gamma$ is the drag coefficient and $\bm\xi$ is the
random force. It is assumed that the latter 
is a Gaussian random variable that
satisfies
the conditions
\begin{equation}
	\langle \xi_i(t)\rangle =0 
	\label{random1}
\end{equation}
and
\begin{equation}
	\langle \xi_i(t_1) \xi_j(t_2)\rangle = 2\mathcal{D}
	\delta_{ij}
	 \delta(t_1 - t_2),
	\label{random2}
\end{equation}
where $\mathcal{D}$ is the diffusion coefficient.
The assumption in Eq.~\eqref{random2} is that 
the motion is a Markovian process, namely that
no correlation of the random force at two different times
exists.

For what follows it is useful to consider the 
momentum broadening, that for a one-dimensional motion reads
$\sigma_p = \langle
(p- p_0)^2\rangle$; in the case of the
Brownian motion we have \cite{Liu:2020cpj}
\begin{equation}
	\sigma_p = p^2_0(e^{-\gamma t}-1)^2 + \frac{\mathcal{D}}{\gamma}(1- e^{-2\gamma t}).
	\label{e26}
\end{equation}	
For times smaller than the equilibration time, $t\ll 1/\gamma$,
we find $\sigma_p \approx 2\mathcal{D} t$.

\section{Results}

In our calculations, we fix the saturation scale, $Q_s$,
and the QCD coupling, $g$. For the latter we use 
the one-loop QCD $\alpha_s$ at the scale $Q_s$; then, we compute
$g=\sqrt{4\pi\alpha_s}$. Therefore
we have $g=g(Q_s)$. From these two
parameters we fix $g\mu$, to be used in Eq.~\eqref{eq:colorchag88},
by virtue of $g\mu \approx Q_s/0.6 g$ \cite{Lappi:2007ku}.
For each set of parameters, we perform $N_\mathrm{events}$
initializations and evolutions up to $t=1$ fm/c, then average
physical quantities over all the events.
Unless specified differently, all the results correspond to
$N_\mathrm{events}=150$, having verified that this is enough
to achieve convergence.

\subsection{Momentum broadening in the
static box}

 \begin{figure}[t!]
	\centering
	\includegraphics[scale=.3]{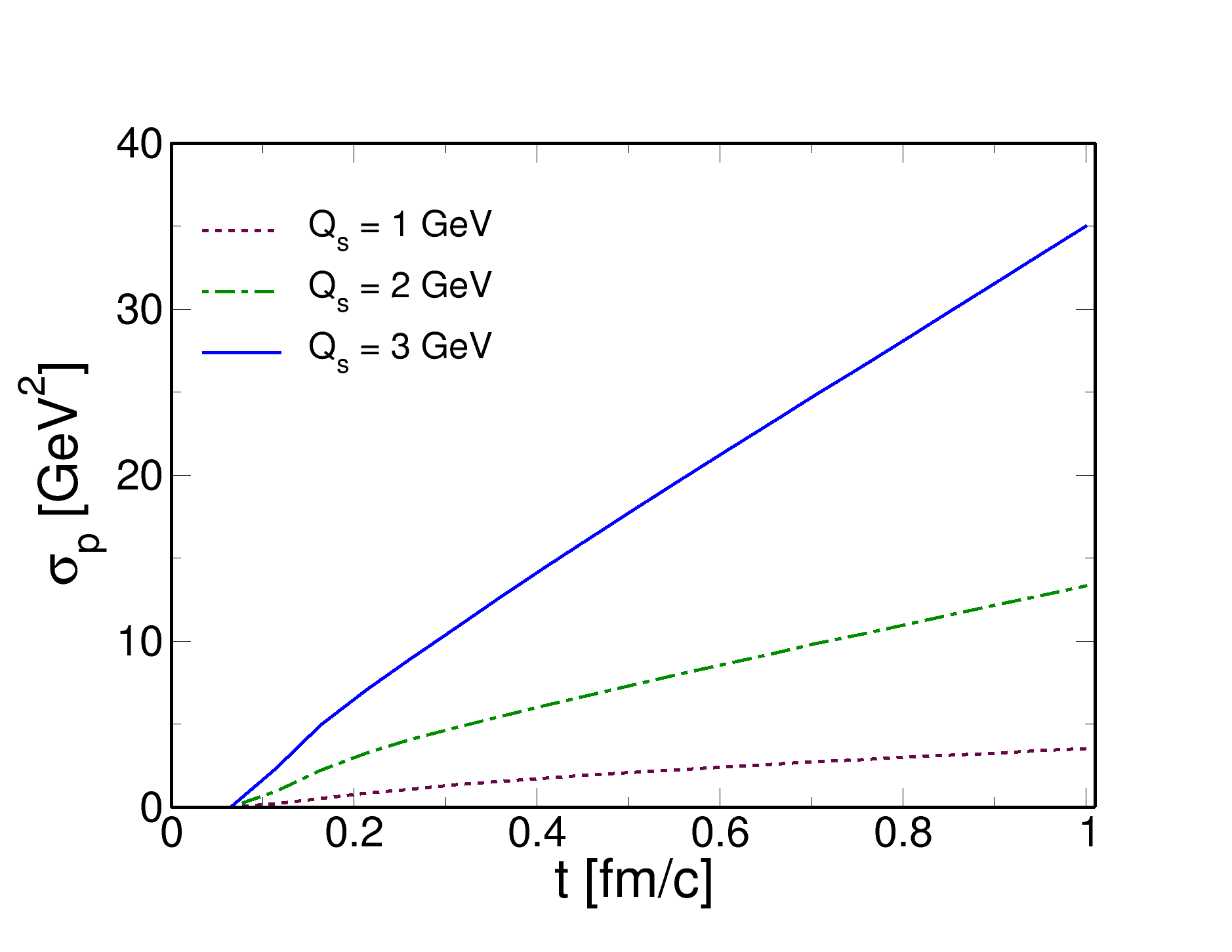}\\
	\includegraphics[scale=.3]{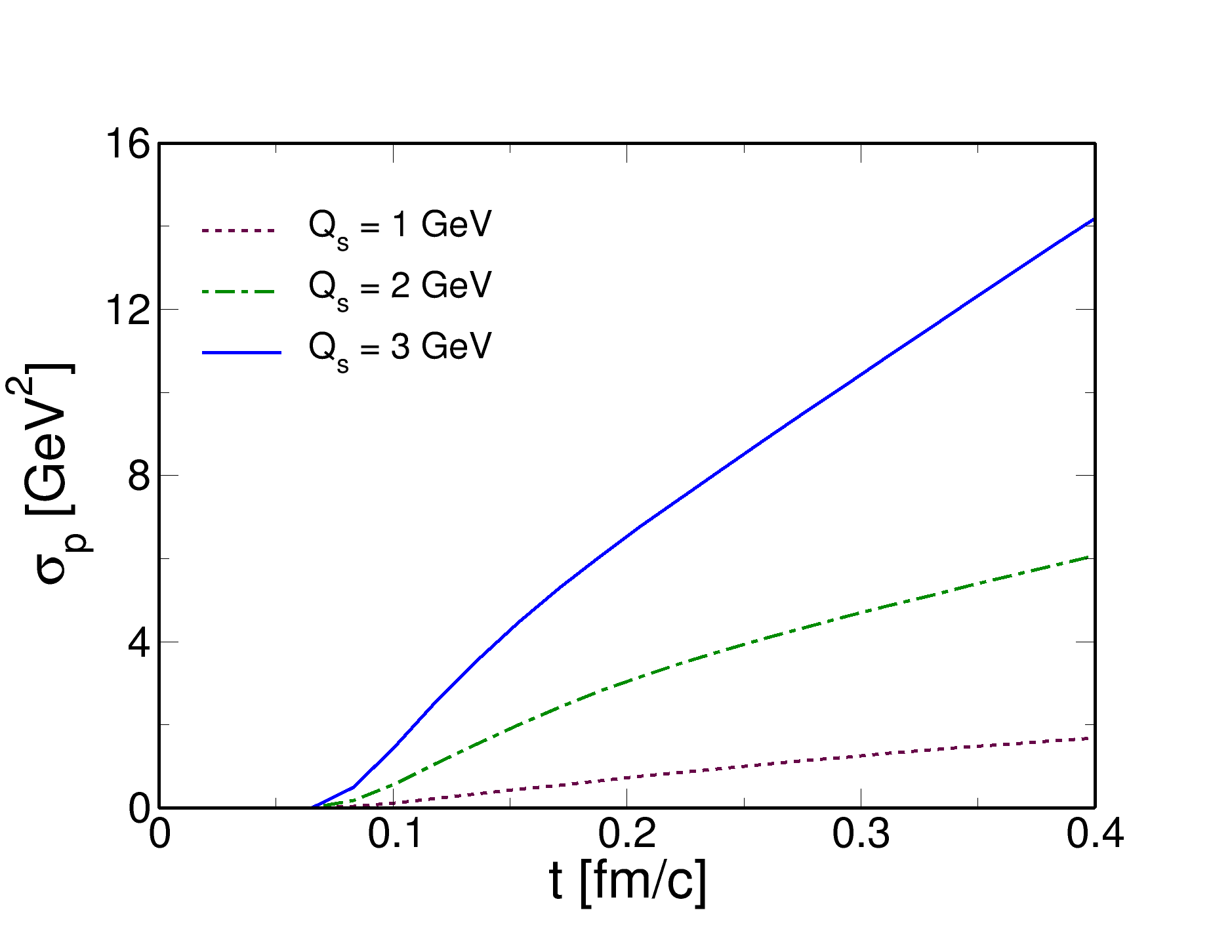}
	\caption{$\sigma_p$ versus proper time for charm quarks,
for the		initial values $p_T=0.5$ GeV.
	The calculation corresponds to evolving Glasma fields
in a static box.}
	\label{Fig:sigmaP}
\end{figure}

In Fig.~\ref{Fig:sigmaP} we plot $\sigma_p$,
defined in Eq.~\eqref{e32}, for charm quarks
versus proper time, for several values of $Q_s$: tuning
this parameter results in the change of the energy density,
thus of the effective temperature, of the Ev-Glasma medium in which
heavy quarks diffuse.
In the lower panel of Fig.~\ref{Fig:sigmaP} we zoom to 
the very early stage, up to $\tau=0.4$ fm/c that roughly
corresponds to the time range in which the Ev-Glasma is 
relevant in realistic collisions.
Calculations correspond to the initial value $p_T=0.5$ GeV;
results for different initializations are similar. 

It is interesting that in the very early time,
$\sigma_p$ versus time is not linear, 
differently from the standard Brownian motion, see 
\eqref{e26}. 
As analyzed in  \cite{Liu:2020cpj},
this difference 
is due to the correlations
of the Lorentz force acting on the charm quarks 
at different times, namely to the memory of the gluon fields.
In \cite{Liu:2020cpj} the memory argument has been
supported by a calculation of the color-electric field
in the EvGlasma: a direct calculation of the gauge-invariant
correlator of the Lorentz force acting on the heavy quarks
confirms this scenario \cite{Dana:Private}.
The memory time, $\tau_\mathrm{mem}$, has been estimated in 
\cite{Liu:2020cpj} by means of the decay of the correlator
of the electric field in the EvGlasma, 
and it has been found $\tau_\mathrm{mem}\approx 1/Q_s$.
After the initial transient, $\sigma_p$ grows up linearly
as in the Brownian motion without a drag \cite{Liu:2020cpj}.
The shift among the two regimes happens on a time scale
$t\approx\tau_\mathrm{mem}$.

\begin{figure}[t!]
	\centering
	\includegraphics[scale=.3]{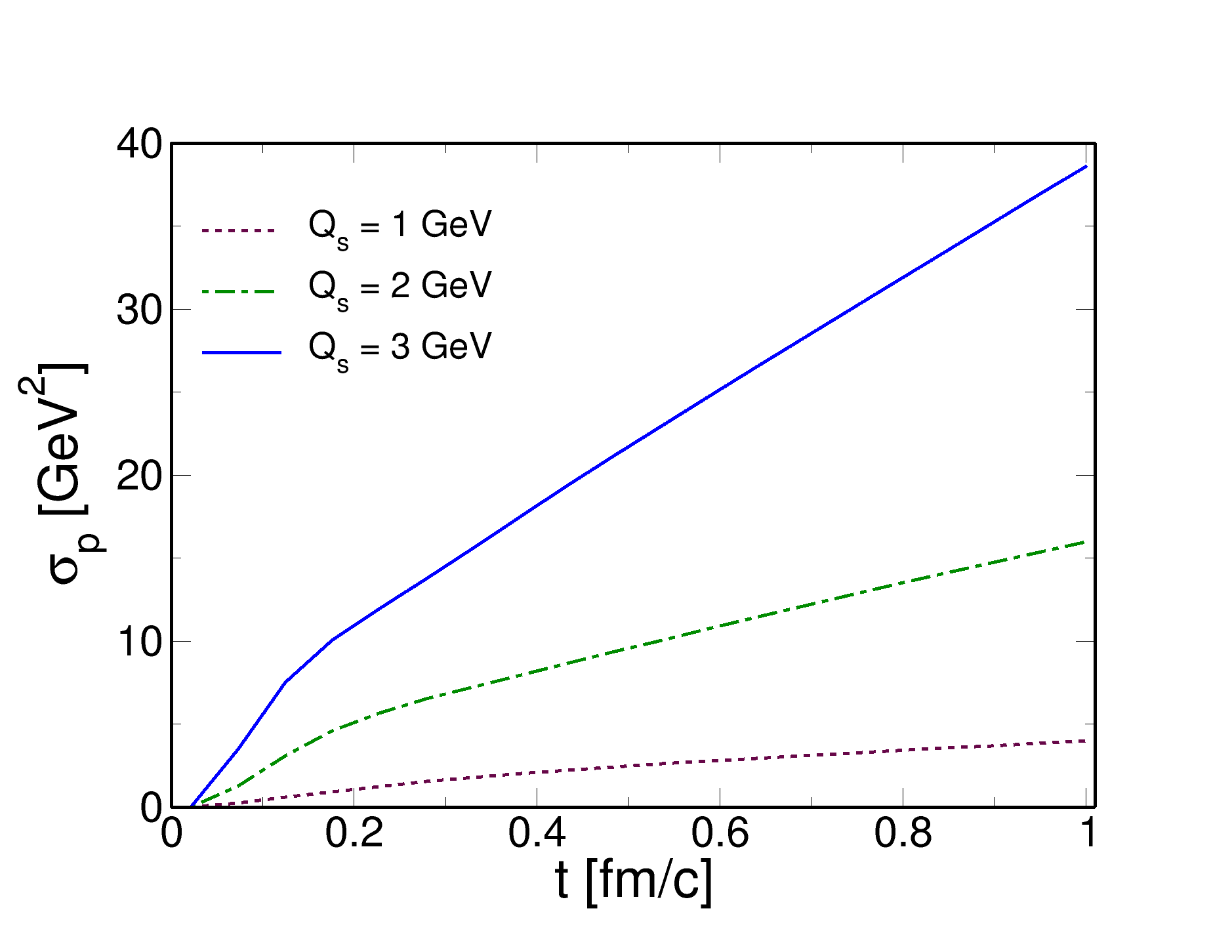}\\
	\includegraphics[scale=.3]{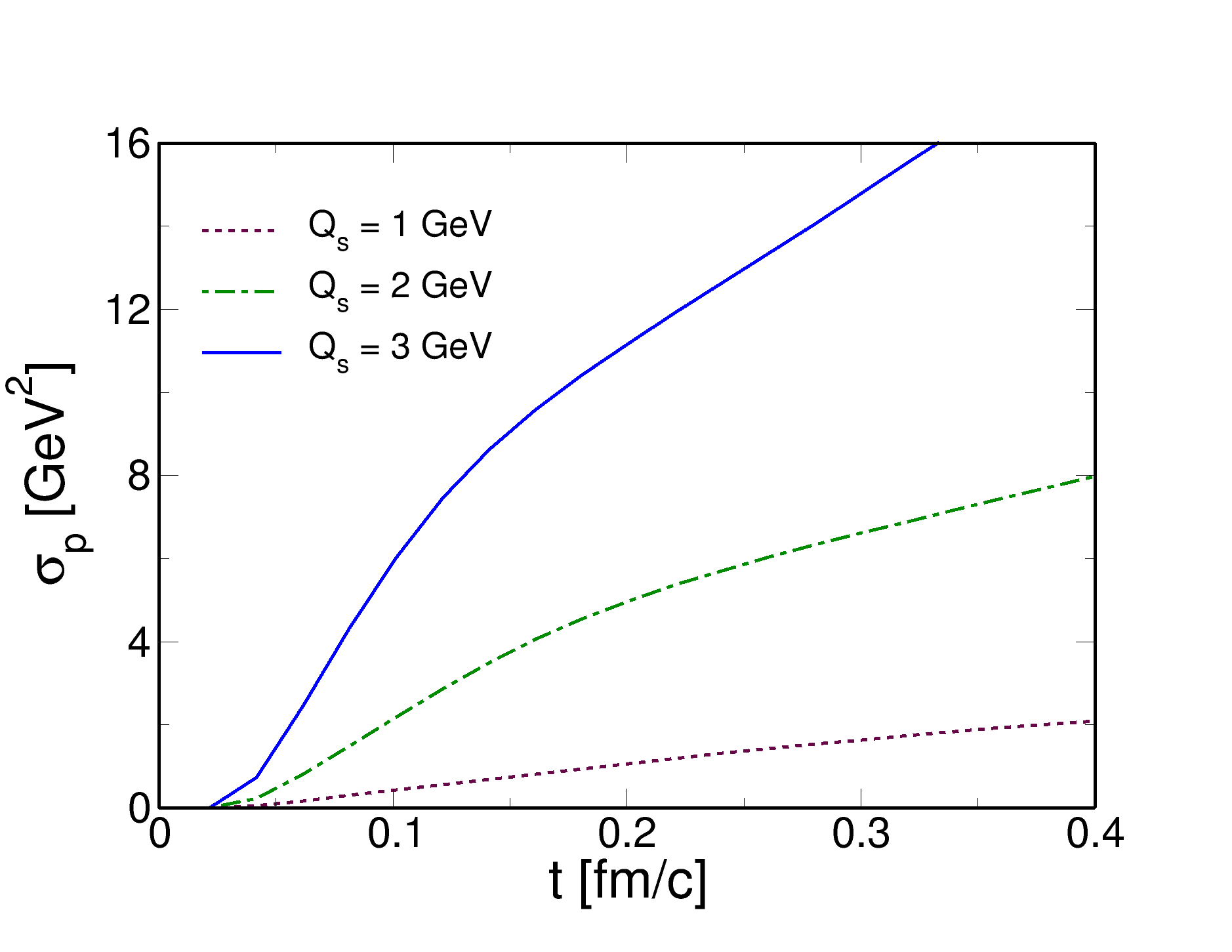}
	\caption{$\sigma_p$ versus proper time for beauty quarks.
		Initial values of $p_T=0.5$ GeV.
		The calculation corresponds to evolving Glasma fields
		in a static box.}
	\label{Fig:sigmaP_run}
\end{figure}

In Fig.~\ref{Fig:sigmaP_run} we plot $\sigma_p$ for 
beauty quarks; the results agree qualitatively 
with those obtained for the charm quarks.
The only relevant difference that we have found is that 
$\sigma_p$ of beauty quarks is slightly larger than that of
charms. 

\begin{figure}[t!]
	\centering
	\includegraphics[scale=.3]{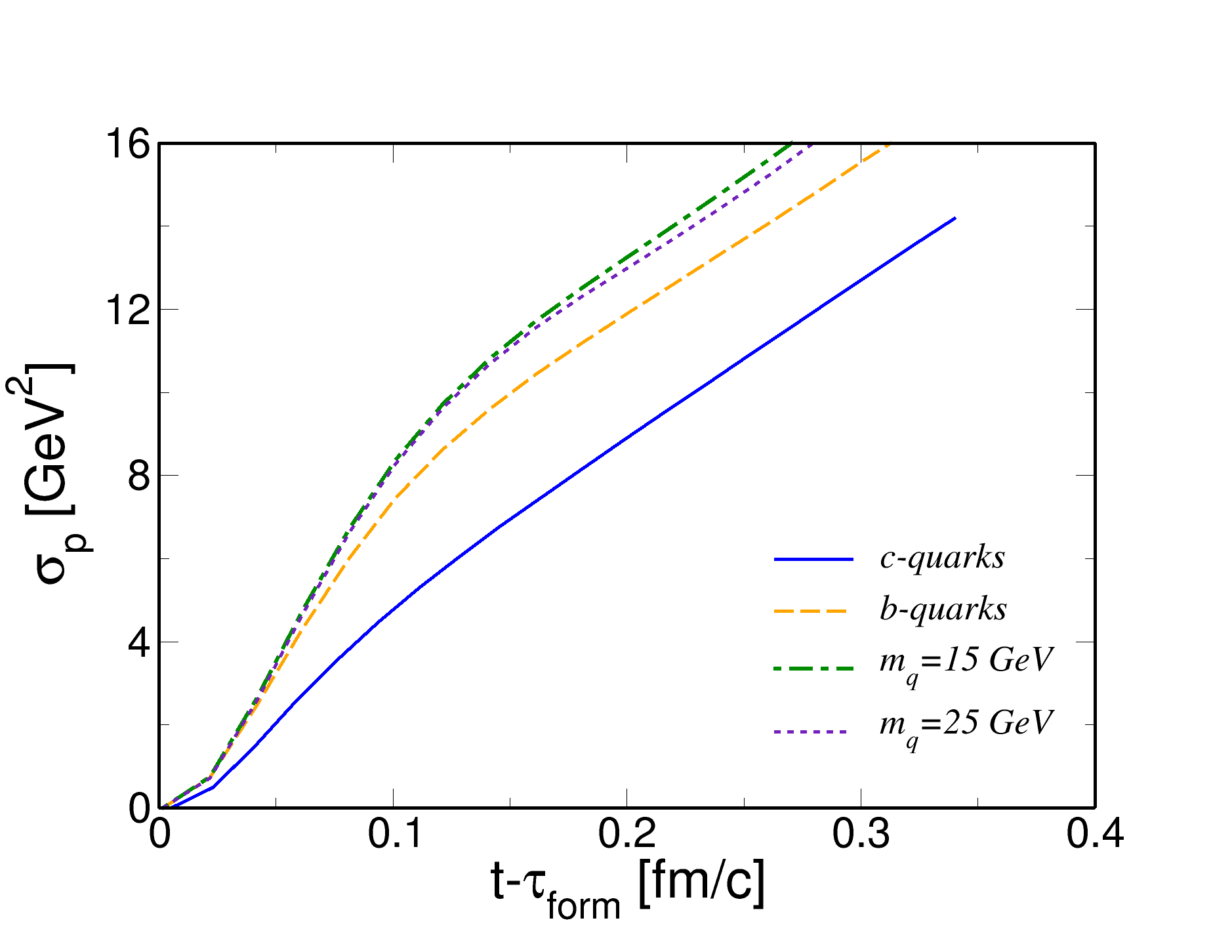}
	\caption{ 
		$\sigma_p$ versus time for several values of
		the heavy quark mass and for
	$Q_s=3$ GeV.}
	\label{Fig:sigmaP_COMP}
\end{figure}

In order to explore this difference in more detail,
in Fig.~\ref{Fig:sigmaP_COMP} we plot 
$\sigma_p$ for $Q_s=3$ GeV and for several
values of the
heavy quark mass; we limit to show results up to $0.4$ fm/c
since after that time momentum diffusion is already in the
linear regime and nothing exciting happens.
In the figure, we have put $\tau_\mathrm{form}=0.02$ fm/c
for all quarks,
that is the initialization time of beauty.
We find that increasing the heavy quark mass
results in the increase of $\sigma_p$ and in its faster
evolution. This can be understood easily: 
on average, heavy quarks with larger mass and same $p_T$
have smaller transverse velocity, therefore they spend more time
within one correlation domain (that is,
the same flux tube) of the EvGlasma and are 
accelerated more efficiently by the gluon field,
resulting in $\sigma_p\propto t^2$;
then their speed becomes large enough that they can move within
uncorrelated domains and experience a Brownian motion, 
resulting in the linear increase of $\sigma_p$.

\begin{figure}[t!]
	\centering
	\includegraphics[scale=.3]{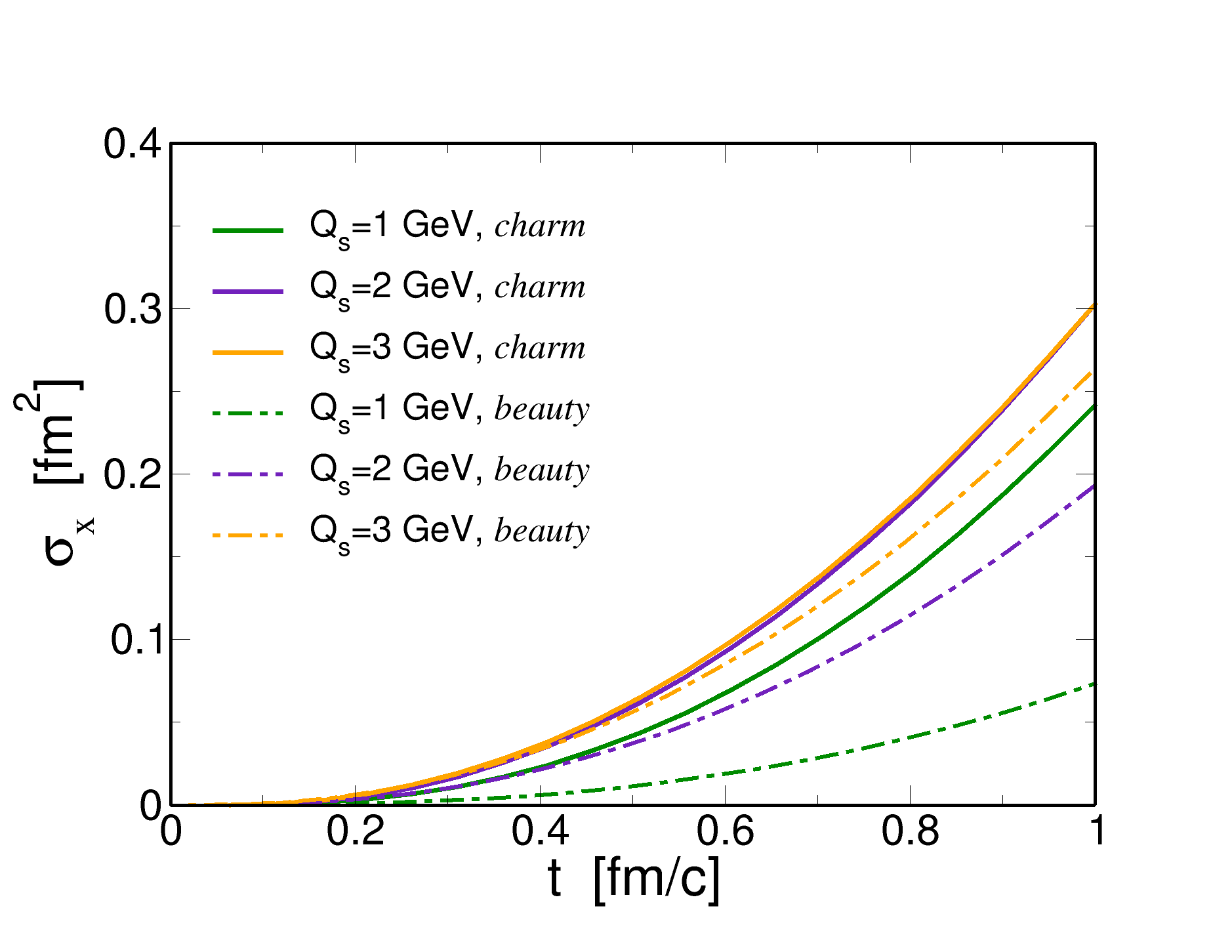}
	\caption{
		$\sigma_x$ versus time for charm quarks (solid lines)
		and beauty quarks (dot-dashed lines).
		}
	\label{Fig:sigmaX_both}
\end{figure}

This picture is supported by the spatial diffusion in the
transverse plane. In Fig. \ref{Fig:sigmaX_both} we plot
$\sigma_x$,
	\begin{equation}
	\sigma_x = \frac{1}{2} \left\langle 
	(x(t)-X_{0})^2 + (y(t)-Y_{0})^2 \right\rangle,
	\label{e3aaa2}
\end{equation}
where $X_{0}$ and $Y_{0}$ are the initial 
transverse plane coordinates of the heavy quark.
Besides, the results in 
Fig. \ref{Fig:sigmaX_both} show that the average velocity of 
beauty quarks is smaller than that of charm quarks, as the formers
spread more slowly in the transverse plane. Therefore,
beauty quarks spend more time within a correlation domain
and get accelerated coherently by the gluon fields.

\begin{figure}[t!]
	\centering
	\includegraphics[scale=.3]{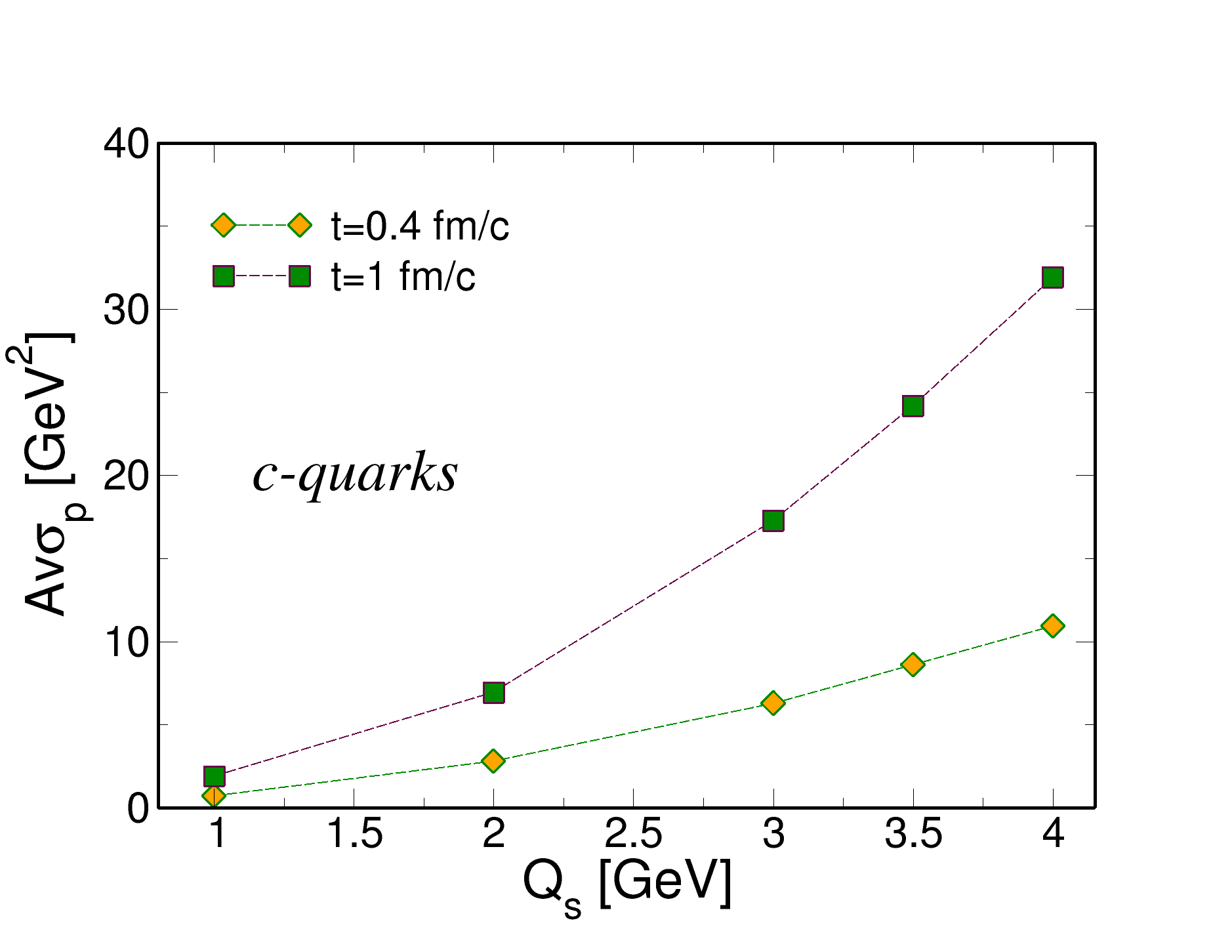}\\
	\includegraphics[scale=.3]{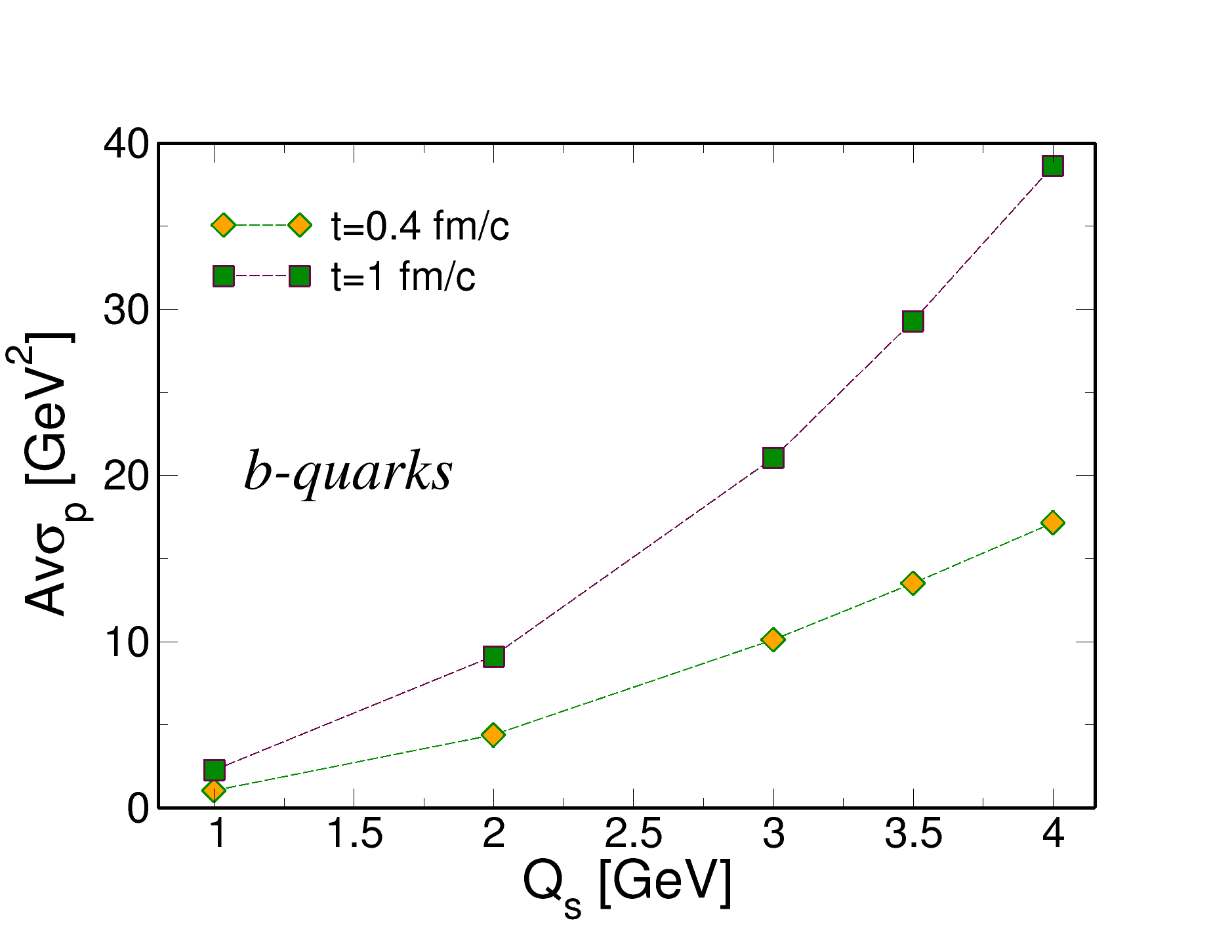}
	\caption{Time average of 
		$\sigma_p$ versus $Q_s$ for charm quarks (upper panel)
		and beauty quarks (lower panel).
Average has been computed on the time ranges $t=0.4$ fm/c
and $t=1$ fm/c respectively.}
	\label{Fig:sigmaP_ave}
\end{figure}

In Fig.~\ref{Fig:sigmaP_ave} we plot the time averaged
$\sigma_p$ versus $Q_s$, 
\begin{equation}
\mathrm{Av}\sigma_p = \frac{1}{t}\int_0^t\sigma_p(t^\prime)
dt^\prime.
\label{eq:delAVsp}
\end{equation}
This quantity is of some interest because it measures
the total momentum spreading of the HQs, therefore it is useful
in the comparison with the Langevin diffusion, see below.
We have computed the average  
for $t=0.4$ fm/c (orange diamonds) and
$t=1$ fm/c (green squares). 
In the upper panel we show
the results for the charm quarks, while in the lower panel
we plot those for the beauty quarks. 
Comparing the results of charm and beauty,
again we notice
the slightly higher values obtained for the latter.

\subsection{Comparison with the Langevin dynamics
\label{sec:forqs}}

Next we turn to the
comparison of momentum diffusion in the EvGlasma
with those of a standard, collisional Langevin dynamics.
We prepare a bulk of thermalized massless gluons
with the same energy density, $\varepsilon$, of the EvGlasma,
putting
\begin{equation}
\varepsilon = 
2(N_c^2-1)\int \frac{d^3p}{(2\pi)^3}  \frac{p}{e^{\beta p}-1}
=\frac{(N_c^2-1)\pi^2 T^4}{15},
\label{eq:enedensUUU}
\end{equation}
where $T=1/\beta$ is the temperature of the gluon system. We use \eqref{eq:enedensUUU} to
compute $T$ for a given $\varepsilon$. 
Then, we run simulations of the
Langevin equation with the pQCD diffusion coefficient computed
at the temperature $T$ with the coupling $g=g(Q_s)$ used for the
EvGlasma calculation. 
We show for completeness the $T$ versus $Q_s$ in Appendix \ref{sec:temperature}.
As expected we find $T=O(Q_s)$.

In the Langevin equation \eqref{l2}  and in the
thermal noise average \eqref{random2} we use
the kinetic coefficients of  
$c \ell \rightarrow c \ell$ and $b \ell \rightarrow b \ell$,
where $\ell$ denotes a massless gluon in the thermal medium.
The basic equations for the drag and diffusion coefficients
are well known and can be found in the literature, see
for example \cite{comb, Svetitsky:1987gq}.
Here it is enough to quote the expression of 
the squared invariant scattering amplitude;
for example, for the  $c \ell \rightarrow c \ell$ scattering this
is given by
\begin{eqnarray}
	|{\cal M}_{c\ell  \rightarrow c\ell }|^2={\pi^2 \alpha_s^2} 
	\bigg\lbrack 
	\frac{32(s-m^2)(m^2-u)}{(t-m_D^2)^2}&& \nonumber  \\ 
	+\frac{64}{9}\frac{(s-m^2)(m^2-u)+2M^2(s+m^2)}{(s-m^2)^2} \nonumber  \\ 
	+\frac{64}{9}\frac{(s-m^2)(m^2-u)+2m^2(m^2+u)}{(m^2-u)^2}\nonumber\\
	+ \frac{16}{9}\frac{m^2(4m^2-t)}{(s-m^2)(m^2-4)} \nonumber  \\   
	+16\frac{(s-m^2)(m^2-u)+m^2(s-u)}{(t-m_D^2)(s-m^2)}\nonumber\\
	-16\frac{(s-m^2)(m^2-u)-m^2(s-u)}{(t-m_D^2)(m^2-u)} 
	\bigg\rbrack, 
	\label{eq:Msquared_cgcg}
\end{eqnarray}
where $s$, $t$ and $u$ denote the standard 
Mandelstam variables and $m$ is the mass of the charm quark. 
A similar expression holds for the beauty.
$m_D$ is the Debye mass, that we assume to be
\begin{equation}
m_D = g T.
\label{eq:alpj_ooo_3434}
\end{equation}

We aim to compare the diffusion within the Langevin dynamics
with that in the EvGlasma with the same energy density.
To this end, for each EvGlasma calculation with a $Q_s$
we compute the temperature of the thermalized medium 
by means of \eqref{eq:enedensUUU}.
This temperature enters 
in the Debye mass \eqref{eq:alpj_ooo_3434} 
and in the distribution functions
of the thermalized gluon medium.
Moreover, 
we use the same $\alpha_s=\alpha_s(Q_s)$ 
in both the Langevin and the EvGlasma calculations: 
this enters
both as an overall factor in 
$|{\cal M}_{c\ell  \rightarrow c\ell }|^2$ and 
in the Debye mass.

\begin{figure}[t!]
	\centering
	\includegraphics[scale=.3]{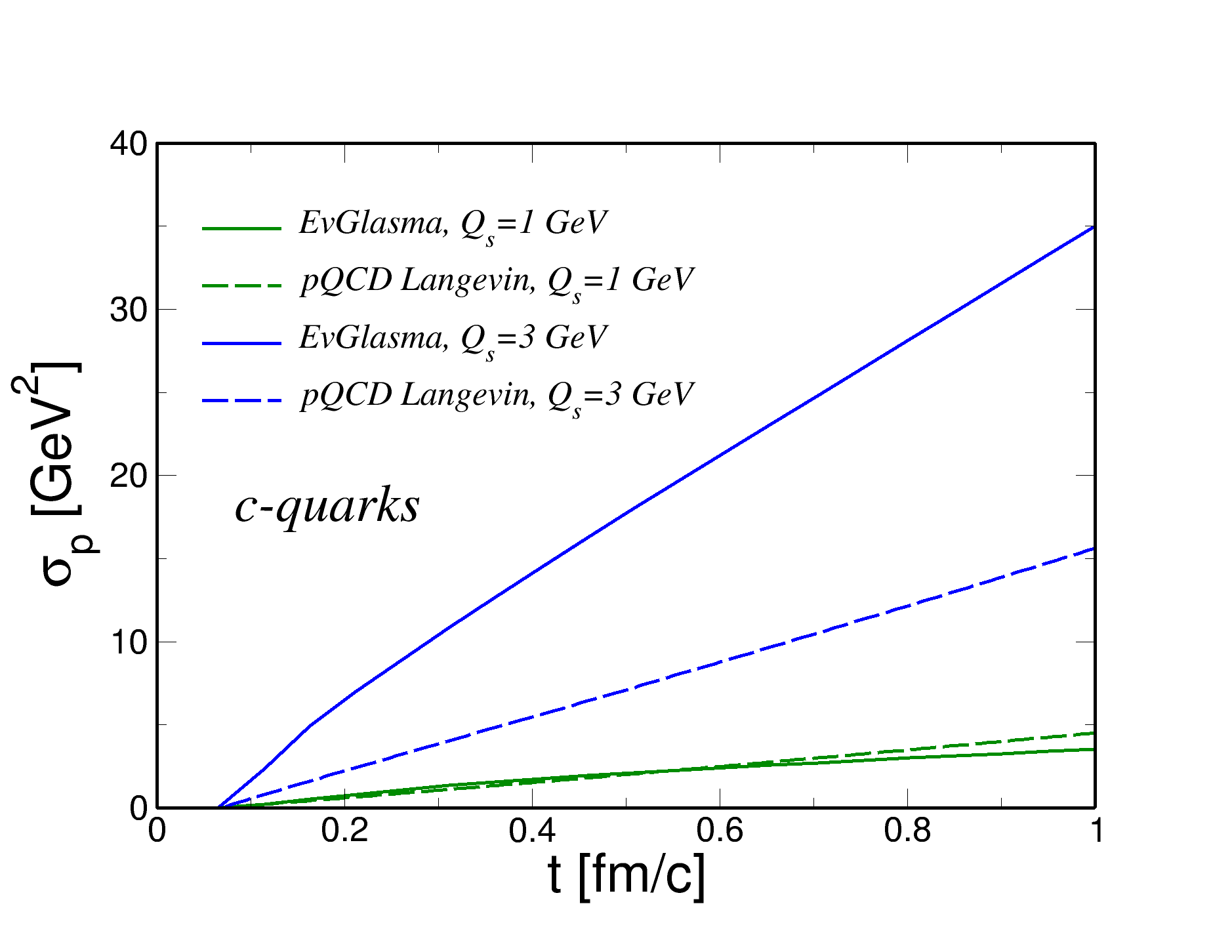}\\
	\includegraphics[scale=.3]{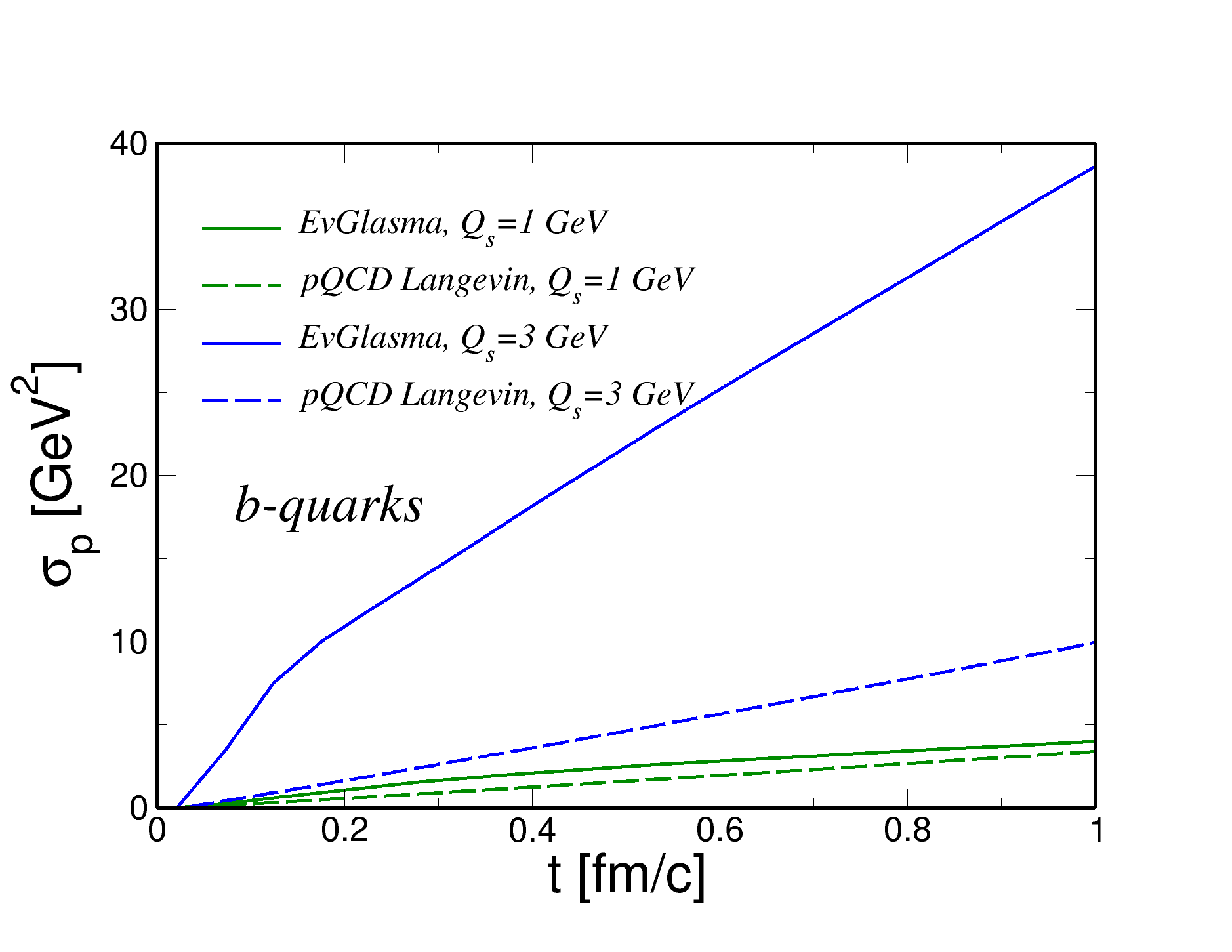}
	\caption{ 
		$\sigma_p$  
		versus time for EvGlasma and pQCD Langevin
		dynamics. The $Q_s$ in the legend for
		each Langevin calculation reminds that 
		the calculation has been performed
	on the top of a gluon bulk with the energy density of the 
	EvGlasma with that $Q_s$.}
	\label{Fig:spcharm}
\end{figure}

\begin{figure}[t!]
	\centering
	\includegraphics[scale=.3]{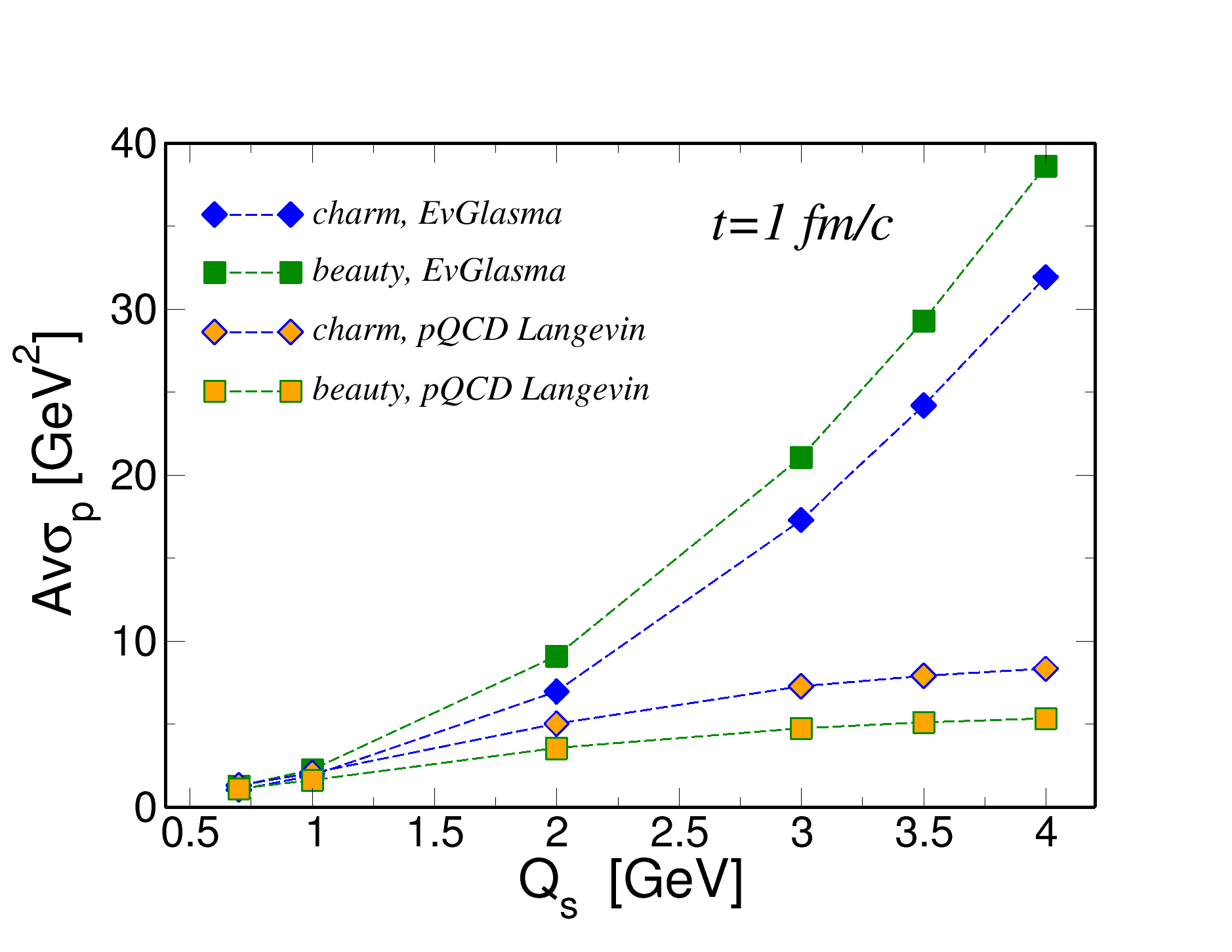}\\
	\includegraphics[scale=.3]{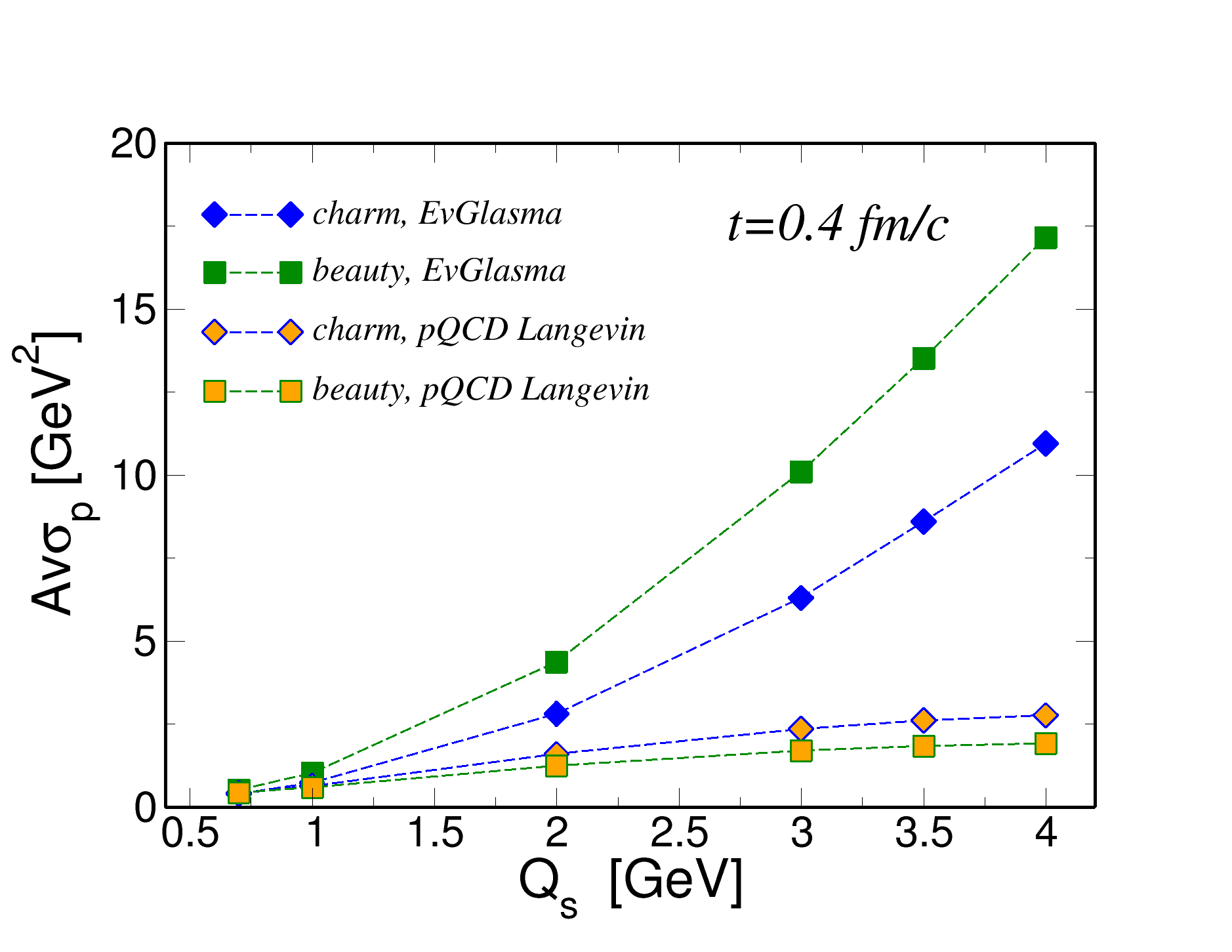}
	\caption{Time average of 
		$\sigma_p$ at $t=1$ fm/c (upper panel)
		and $t=0.4$ fm/c (lower panel)
		versus $Q_s$ for EvGlasma and pQCD Langevin
		dynamics. Calculations correspond to the 
	static box geometry.}
	\label{Fig:sigmaP_aveQs}
\end{figure}

In Fig.~\ref{Fig:spcharm} we plot
$\sigma_p$  
versus time for EvGlasma and pQCD Langevin
dynamics; results for both charm and beauty quarks
are shown. 
In the legend,  
the $Q_s$ for the
Langevin calculations is a reminder that 
the calculation has been performed
on the top of a gluon bulk with the energy density of the 
EvGlasma with that $Q_s$.
In Fig.~\ref{Fig:sigmaP_aveQs} we plot the time averaged 
$\sigma_p$ at $t=1$ fm/c and $t=0.4$ fm/c
for both EvGlasma and Langevin cases.

The Langevin results 
in Figg.~\ref{Fig:spcharm} and~\ref{Fig:sigmaP_aveQs}
are quite interesting.
Firstly, we notice that $\sigma_p$ versus time
is qualitatively different from that
of the EvGlasma because in the former there is no memory,
therefore $\sigma_p\propto t$ in the time range considered
(drag would make the growth of $\sigma_p$ slower for times of the order of
 the thermalization
time of the heavy quarks).

However, looking in particular at Fig.~\ref{Fig:sigmaP_aveQs}
we note that for small $Q_s$, namely when the fields of the
EvGlasma are not intense, the average $\sigma_p$ in the two
calculations is comparable. This is due to the fact that
for small $Q_s$ the energy density is small so the EvGlasma
is closer to a dilute gluon system and momentum diffusion
is similar to that of a collisional dynamics.
For large $Q_s$ the discrepancy between the two systems
becomes substantial, and it is due to the fact that
heavy quarks in the EvGlasma experience strong coherent
gluon fields while the dynamics remains collisional in the
Langevin case. Therefore, one way to read 
Fig.~\ref{Fig:sigmaP_aveQs} is that on average,
the diffusion coefficient of EvGlasma and pQCD are 
similar for small $Q_s$, in the sense that on the same
time scale the momentum spreading is comparable in the
two cases.
  
One interesting point in Fig.~\ref{Fig:sigmaP_aveQs} 
is that in the Langevin case, $\sigma_p$ decreases with
the quark mass, while the opposite behavior is found
for the EvGlasma. This qualitative difference 
can be understood
easily.  In the pQCD Langevin the cross section 
of heavy quarks with the thermal bath is $\propto 1/m$ in the
nonrelativistic limit. On the other hand, 
in the EvGlasma case the larger mass of the quark implies
a smaller velocity in the transverse plane, hence a larger
time spent within a flux tube where the gluon field is
coherent: therefore, beauty quarks experience 
the nonlinear growth $\sigma_p\propto t^2$ for a longer time
than charm quarks do, eventually leading to larger $\sigma_p$.

\subsection{Momentum broadening for the system with the 
longitudinal expansion} 	

In this subsection, we report on the momentum broadening
of charm quarks obtained in a gluon system subject to a 
boost-invariant longitudinal expansion, and we compare it
with that of the gluon system in a static box.
Again, we show some results up to $\tau=1$ fm/c for illustrative
purposes only, keeping in mind that in realistic 
collisions at the LHC the expanding
Ev-Glasma is relevant up to $\tau\approx 0.4$ fm/c. 

In the case of the longitudinally expanding medium it is
convenient to shift from $(t,z)$ 
to $(\tau,\eta)$ coordinates, to take advantage of the
boost invariance of the EvGlasma fields. In this case,
the initial longitudinal, $\eta-$components of the
fields are given by
\begin{eqnarray}
	E^\eta &=& -ig \sum_{i=x,y} \Big[\alpha^{(B)}_i, \alpha^{(A)}_i\Big],\\
	B^\eta &=& -ig \Big(\Big[\alpha^{(B)}_x, \alpha^{(A)}_y\Big]+\Big[\alpha^{(A)}_x, \alpha^{(B)}_y\Big]\Big),
\end{eqnarray}	
while the transverse components vanish;
the $\alpha_i$ fields are defined as in \eqref{eq:alpha1pp}
and \eqref{eq:alpha2pp}.
The CYM equations read
\begin{eqnarray}
	E^{i}&=&\tau \partial_{\tau} A_{i},\\
	E^{\eta}&=&\frac{1}{\tau} \partial_{\tau} A_{\eta},
\end{eqnarray}
and
\begin{eqnarray}
	\partial_{\tau} E^{i} &=&\frac{1}{\tau} D_{\eta} F_{\eta i}+\tau D_{j} F_{j i},\\
	\partial_{\tau} E^{\eta}&=&\frac{1}{\tau} D_{j} F_{j \eta}.
\end{eqnarray}
Finally, the  energy density is given by
\begin{equation}
	\varepsilon = \mathrm{Tr}\left[ \frac{1}{\tau^2}E^i E^i + E^\eta E^\eta +\frac{1}{\tau^2} B^i B^i + B^\eta B^\eta \right].
\end{equation}

\begin{figure}[t!]
	\centering
	\includegraphics[scale=.3]{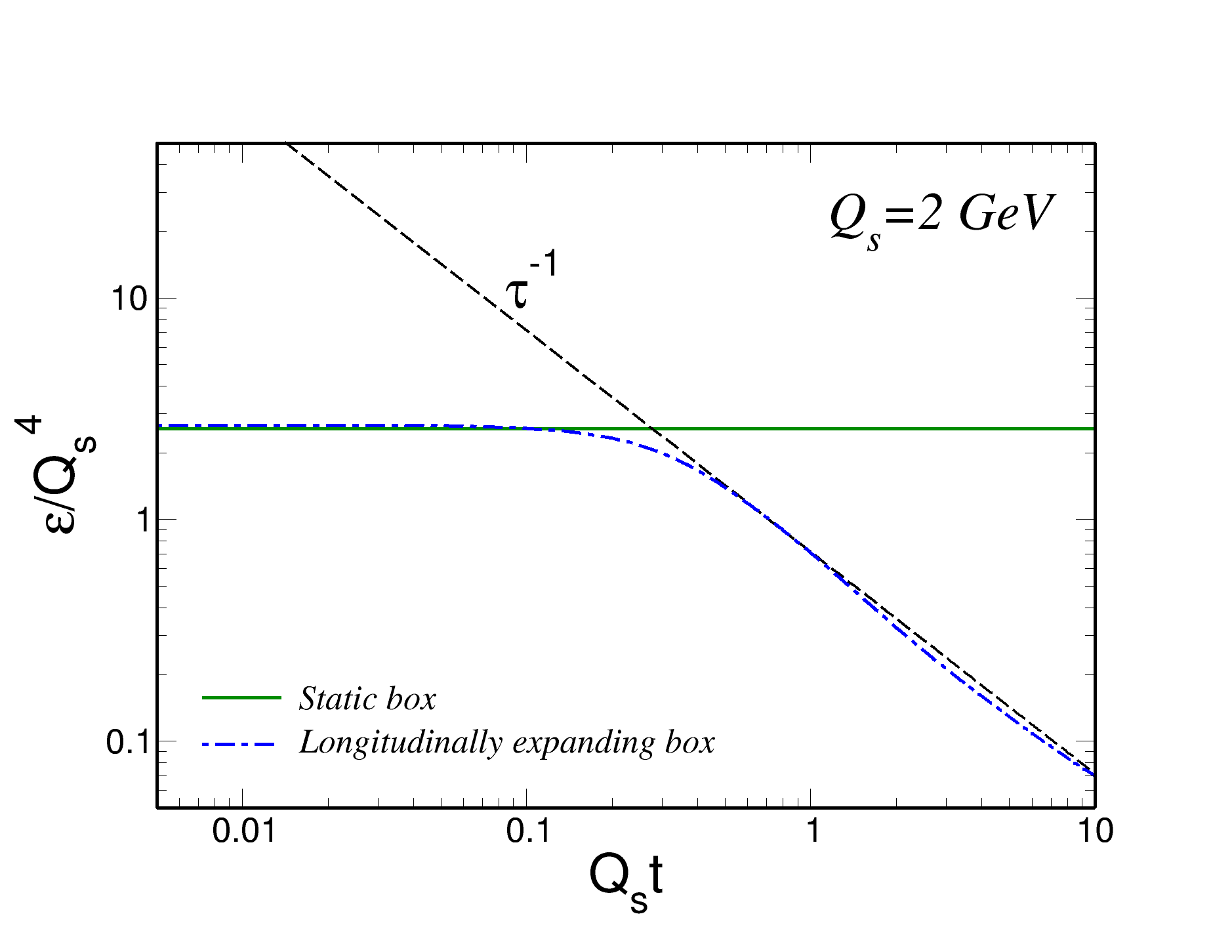}
	\caption{Energy density, $\varepsilon$, versus proper time
	for the static (solid green) and the
longitudinally expanding (dot-dashed blue) boxes.
Calculations correspond to $Q_s=2$ GeV.
Dashed black line guides the eyes on the 
free streaming $\tau^{-1}$ behavior
of $\varepsilon$, that occurs for $Q_s t\approx 1$ namely for
$t\approx0.1$ fm/c.}
	\label{Fig:enedens}
\end{figure}

In Fig.~\ref{Fig:enedens} we plot, for the sake of reference,
the energy density versus the proper time for $Q_s=2$ GeV,
in the cases of a static box and a longitudinally expanding
medium. Obviously $\varepsilon$ stays constant in the 
case of the static box, while it decays with the
free streaming behavior $\varepsilon\propto\tau^{-1}$ 
after a short
transient in the case of the expanding system.
The transition from the initial state to the free streaming
solution happens around $Q_s t\approx 1$.

 \begin{figure}[t!]
 	\centering
 	\includegraphics[scale=.3]{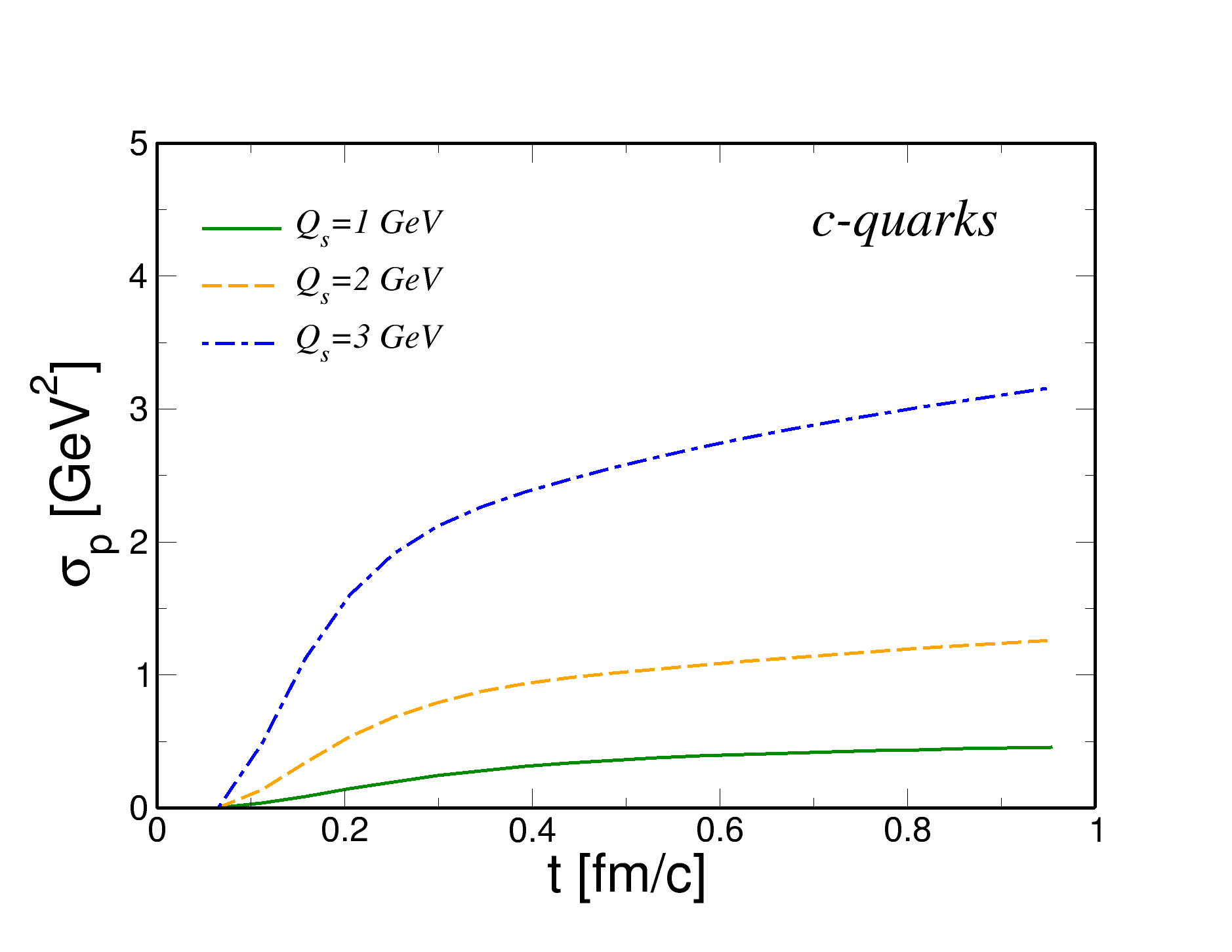}\\
 	\includegraphics[scale=.3]{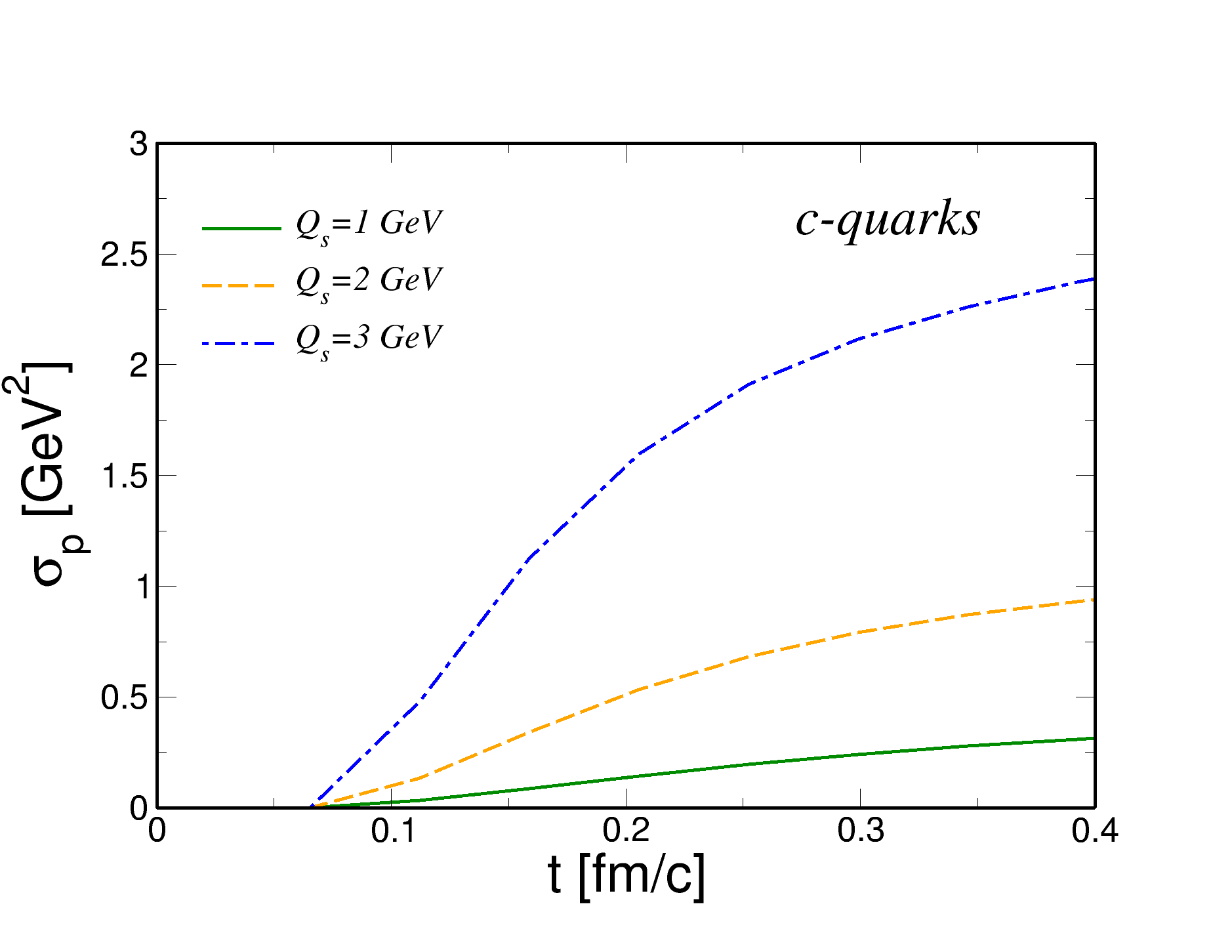}
 	\caption{$\sigma_p$ versus proper time for charm quarks.
 Values of $Q_s$ used in the calculations are shown in the
legend. In the upper panel we perform calculations
up to $\tau=1$ fm/c while in the lower panel we zoom in
the range up to $\tau=0.4$ fm/c. }
 	\label{Fig:comparison}
 \end{figure}

In Fig.~\ref{Fig:comparison} we plot $\sigma_p$ of charm quarks,
computed for a 
system subject to a boost-invariant longitudinal expansion,
and we compare it with that obtained for a static
box. The longitudinal expansion mimics better the initial
stage of the high energy nuclear collisions.
In the lower panel of Fig.~\ref{Fig:comparison}  we offer a 
zoom of the upper panel for very early time,
up to $\tau\approx 0.4$ fm/c, which corresponds to the
effective time range in which our calculation is 
phenomenologically interesting.
 	  
\begin{figure}[t!]
	\centering
	\includegraphics[scale=.3]{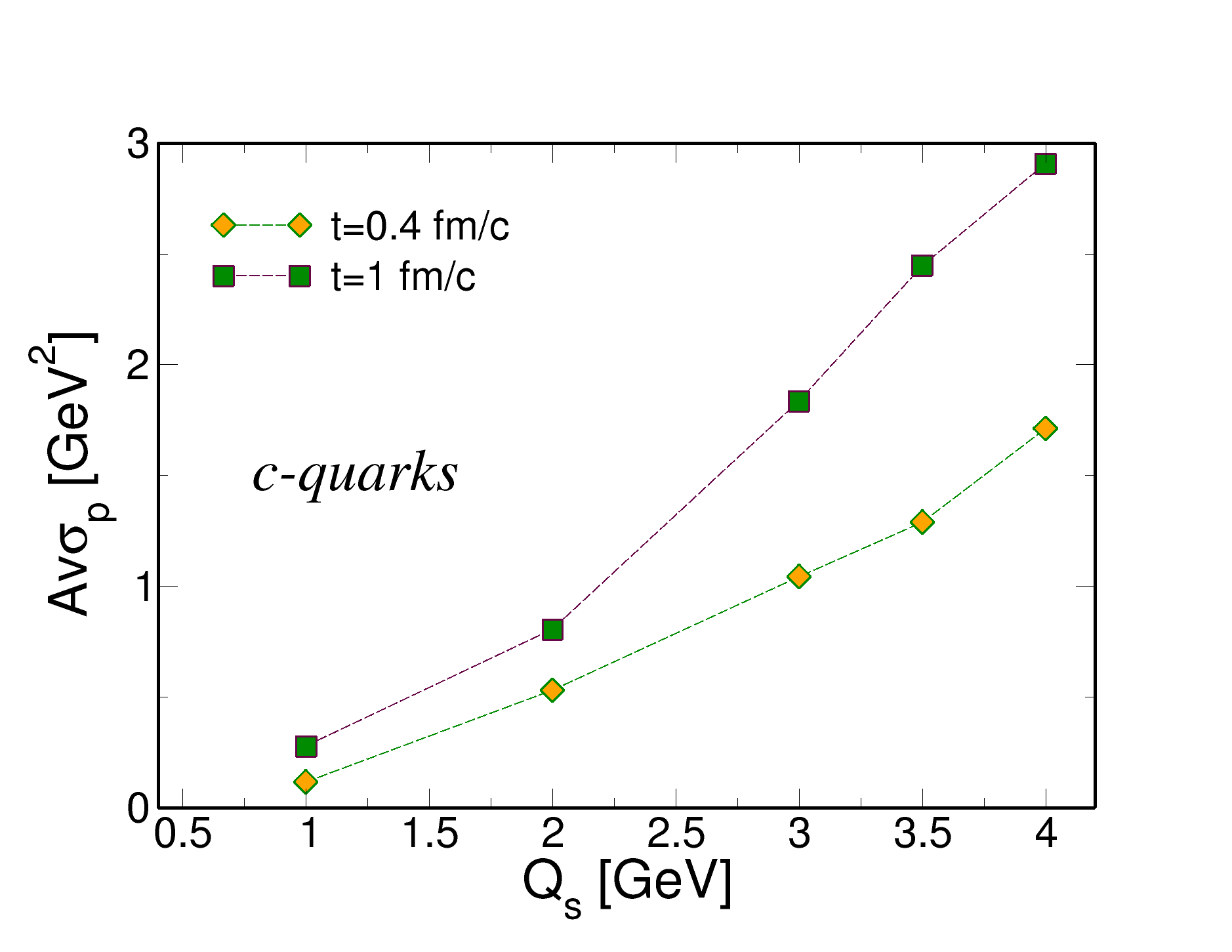}
	\caption{Time average of $\sigma_p$ for charm quarks,
	in the case of the longitudinally expanding box.}
	\label{Fig:expsp}
\end{figure}
 
For the sake of completeness, in  Fig.~\ref{Fig:expsp} we plot
the time average of  
 $\sigma_p$ for charm quarks,
in the case of the longitudinally expanding box; results for 
beauty are similar.
In comparison with the results of the
static box, see Fig.~\ref{Fig:sigmaP_aveQs}, we note that
the expansion lowers the average $\sigma_p$ of a factor
of about $2$ to $5$: this is expected due to the 
dilution of the energy density for the expansion.
However, as shown already in \cite{Liu:2019lac},
this is enough to tilt the spectrum of the heavy quarks,
see also \cite{Sun:2019fud}.
A systematic study of the potential effects on the
observables will be the subject of a future study.

\section{Conclusions}
We have analyzed in some detail the diffusion of heavy quarks,
charm and beauty, in the strong gluon fields that are produced
in the early stage of relativistic heavy ion collisions.
The dynamics of the fields is described by the classical
Yang-Mills equations with the Glasma initialization,
while heavy quarks are evolved via the relativistic kinetic
Wong equations in the probe approximation.
The evolving gluon fields have been baptized as the
EvGlasma, where Ev stands for evolving.

We have studied the transverse momentum broadening, $\sigma_p$,
as well as the diffusion in the transverse coordinate plane,
$\sigma_x$, for both $c$ and $b$. We have found that,
differently from the standard Brownian motion, $\sigma_p\propto
t^2$ in the very early stage, while the linear behavior
$\sigma_p\propto t$ is achieved in the later stage.
The early stage behavior is related to the  
memory in the Lorentz force acting on the heavy quarks;
the linear increase of $\sigma_p$ with time
in the later stage
is interpreted by noticing that heavy quarks get enough 
transverse velocity by the color fields, so on average they can
jump from one flux tube to another: since the tubes 
are placed randomly in the transverse plane, fast heavy quarks
experience, on average, a diffusion in a random medium
like particles in the standard Brownian motion without a drag
force, hence 
$\sigma_p\propto t$.

We have found that momentum diffusion in the EvGlasma
is enhanced by the quark mass: we understand this by noticing that
that the larger the mass the slower the heavy quark is,
therefore a large mass quark spends more time within a single
flux tube than a light quark, and experiences the nonlinear
growth of $\sigma_p$ resulting in a larger averaged $\sigma_p$.
This interpretation is supported by the coordinate
diffusion, that instead is suppressed by the quark mass,
confirming that the large mass quarks diffuse into small
regions of the transverse plane.

We have then compared the diffusion in the EvGlasma with
that of a standard Langevin dynamics: the kinetic coefficients
in the latter case have been computed within pQCD at 
finite temperature. To this end, we have prepared
a bath of thermalized, massless gluons with energy density,
$\varepsilon$, equal to that of the EvGlasma for a given value
of $Q_s$. From $\varepsilon$ we have extracted the temperature,
$T$, that we have used to compute the screened cross sections
in the thermal medium. The strong coupling used in the 
Langevin calculations is the same that we have used for the
EvGlasma, $g=g(Q_s)$.

We have found that for small $Q_s$ the average $\sigma_p$ 
in the two systems is fairly the same. This can be interpreted
by noticing that for small $Q_s$ the EvGlasma is made of
diluted gluon fields, thus the dynamics of the heavy quarks
is expected to be similar to a collisional one.
This result also shows that the diffusion coefficient
of heavy quarks in the EvGlasma is in agreement with
that of pQCD, in the case of small energy density.

On the other hand, quantitative differences appear for
large $Q_s$: in this case, the dynamics of the heavy quarks
in the EvGlasma cannot be reduced to a collisional one
because the gluon fields are too intense. 
In particular, the Langevin dynamics with pQCD coefficients
underestimates the $\sigma_p$, meaning that one would
need to input a larger diffusion coefficient 
in the Langevin equations to reproduce the $\sigma_p$
in the EvGlasma.
As a result,
one cannot use the Langevin equations with pQCD
coefficients to analyze the
diffusion in intense EvGlasma 
and the relativistic kinetic equations
coupled to the evolving gluon fields have to be
considered.

Another interesting result, summarized in Fig.~\ref{Fig:sigmaP_aveQs}, is that beauty quarks are more
affected by the evolution in the EvGlasma: this is understood 
because beauty quarks are slower, then they spend more time
within a single filament and experience more
the coherent gluonic fields that lead to $\sigma_p\propto t^2$. 

We have also studied the effect of the longitudinal expansion
on the transverse momentum broadening. For this we have limited
to a few representative values of $Q_s$ and to the charm
quarks, since results for different $Q_s$ and beauty quarks
are similar. We have found that in this case, $\sigma_p$ 
tends to saturate while for the static box it increases
indefinitely. We can explain this easily in terms of the dilution
of the energy density due to the expansion, that implies that
the average magnitude of the color fields, thus of the
force acting on the heavy quarks, becomes smaller and the
EvGlasma diffuses $p_T$ less efficiently. 

In the future we would like to study in a more systematic way
the impact of the momentum broadening in the early stage 
on some observables: among these, collective flows
and two-particle correlations are of a certain interest,
both in pA and in AA collisions.

We could improve our calculations in several ways. 
Firstly, 
the drag force could be introduced by adding the color current
carried by $c$ and $b$ in the Yang-Mills equations.
This force
would slow down the evolution of $\sigma_p$
with time; however, as shown in \cite{Liu:2020cpj}, 
the
equilibration time of heavy quarks in th EvGlasma is of the order
of $\tau_\mathrm{equil}\approx 10$ fm/c at least, 
and any effect of the drag would become evident only for
$t\approx \tau_\mathrm{equil}$. Therefore, this would not
affect the early evolution in a substantial way;
however, its inclusion is possible and we plan to add it in our
future works. Another improvement will be the formulation
with three colors. Finally, we aim to prepare realistic
initializations that include initial state fluctuations,
the correct geometrical shape of the initial state and 
the small-$x$ evolution of the initial color charges.
We plan to add all these things in future works.

\vspace{2mm}
\section*{Acknowledgments}
The authors acknowledge J.H. Liu for discussions and technical
support with the earlier version of the code. 
M.R. acknowledges Laboratori Nazionali del Sud in Catania
where this work has been completed,
and in particular V. Greco, 
for the hospitality and for the numerous discussions.
M.R is grateful to D. Avramescu for clarifications
about the force correlator in the EvGlasma,
and to John Petrucci for inspiration.
S.K.D. and M.R. acknowledge the support by the National Science Foundation of China (Grants No.11805087 and No. 11875153). S.K.D. acknowledges the support from DAE-BRNS, India, Project No. 57/14/02/2021-BRNS.

\appendix
\section{Temperature\label{sec:temperature}}

\begin{figure}[t!]
	\centering
	\includegraphics[scale=.3]{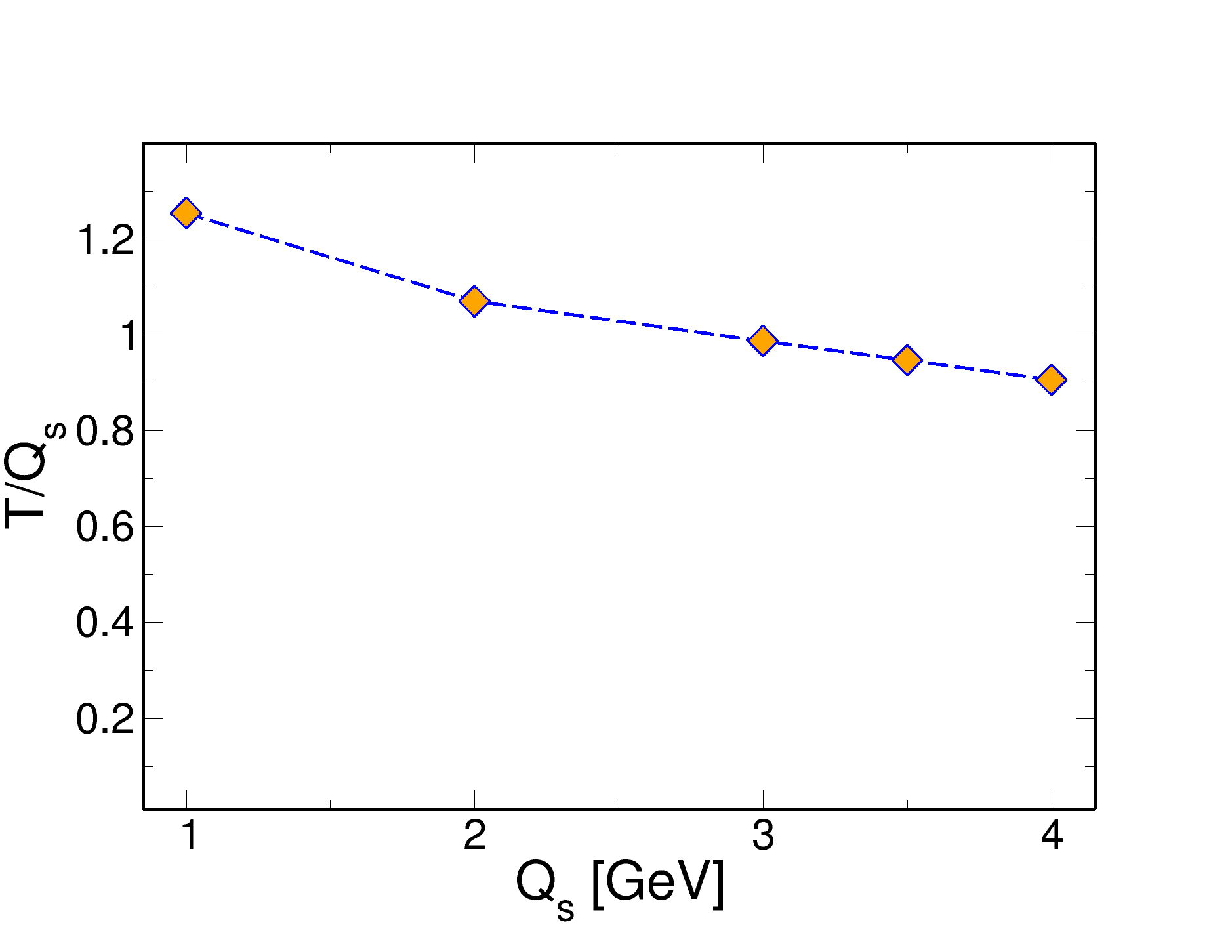}
	\caption{Temperature of a massless gluon system versus
		$Q_s$ obtained by
		means of Eq.~\eqref{eq:enedensUUU}, where $\varepsilon$
		is the energy density of Glasma corresponding
		to the specified
		value of $Q_s$.}
	\label{Fig:tempappendix}
\end{figure}

In Fig.~\ref{Fig:tempappendix} we plot, for completeness,
the temperature of a massless gluon system versus
$Q_s$ obtained by
means of Eq.~\eqref{eq:enedensUUU}, where $\varepsilon$
is the energy density of Glasma corresponding
to the specified
value of $Q_s$. This is the temperature that we use
to compute the diffusion coefficient in pQCD that we have
used in the Langevin equation in section \ref{sec:forqs}.

\end{document}